# A frequency-adjustable electromagnet for hyperthermia measurements on magnetic nanoparticles

L.-M. Lacroix, J. Carrey* and M. Respaud

Université de Toulouse; INSA; UPS; LPCNO, 135 avenue de Rangueil, F-31077 Toulouse, France and
CNRS; LPCNO, F-31077 Toulouse, France

**Abstract :**

We describe a low-cost and simple setup for hyperthermia measurements on colloidal solutions of magnetic nanoparticles (ferrofluids) with a frequency-adjustable magnetic field in the range 5-500 kHz produced by an electromagnet. By optimizing the general conception and each component (nature of the wires, design of the electromagnet), a highly efficient setup is obtained. For instance, in a useful gap of 1.1 cm, a magnetic field of 4.8 mT is generated at 100 kHz and 500 kHz with an output power of 3.4 W and 75 W, respectively. A maximum magnetic field of 30 mT is obtained at 100 kHz. The temperature of the colloidal solution is measured using optical fiber sensors. To remove contributions due to heating of the electromagnet, a differential measurement is used. In this configuration the sensitivity is better than 1.5 mW at 100 kHz and 19.3 mT. This setup allows one to measure weak heating powers on highly diluted colloidal solutions. The hyperthermia characteristics of a solution of Fe nanoparticles are described, where both the magnetic field and the frequency dependence of heating power have been measured.

# Main Text:

## I. Introduction

Hyperthermia consists in rising the temperature of a cancerous tumour to improve the efficiency of chemotherapy. For this purpose, one of the most promising techniques is based on the use of magnetic nanoparticles [1]. Indeed, the magnetic moment switching or the Brownian motion induced by an alternative magnetic field leads to some energy dissipation. In a therapeutic process, magnetic nanoparticles must be first accumulated inside the tumour and then excited by an alternative magnetic field. The applied magnetic field frequency and amplitude values result from a compromise between the efficiency of the heating and the necessity to avoid eddy currents inside the patient. A common criterion is that the magnetic field-frequency product should not exceed $5 \times 10^9$ $Am^{-1}s^{-1}$ [2]. Active research is done to improve the heating power of superparamagnetic nanoparticles, which are good candidate for this application due to their stability in colloidal solutions. The heating power of superparamagnetic nanoparticles is maximal when the relaxation time ($\tau$) of the magnetic moment matches precisely the excitation frequency. Moreover, the heating power increases with the square of their magnetization. Thus, the optimization of the nanoparticles for this application requires a good control of their size, anisotropy, which determines $\tau$, and of their magnetization to increase the heating power.

Experimentally, the heating power value is determined by measuring the temperature rise of a colloidal solution placed in an alternative magnetic field. The magnetic field is in many cases produced by a coil connected to a high power RF generator so that only single-frequency measurements can be performed [1,3,4,5]. However, hyperthermia measurements as a function of frequency are desirable for a further optimization of the colloidal solution since they allow one to determine parameters such as the resonant frequency of the nanoparticles, the distribution of relaxation time, or to check the matching between ac magnetic susceptibility and heating power measurements. The scarce experiments of hyperthermia as a function of frequency reported in the literature are done using home-made power resonator and coils, but no details are given on the experimental setups [6,7]. In this article, we present the set-up we have developed, which is suitable for hyperthermia measurements as a function of frequency. Through the use of an electromagnet, of Litz wires and a home-made resonating transformer, this setup presents the main advantage that a low power is necessary to produce the alternative magnetic field. As a consequence, no cooling system is required for the experiments, and a power amplifier cheaper than the high power RF generator generally used is needed. To improve the sensitivity of the setup, a differential temperature measurement is used, leading to sensitivity better than 1.5 mW in a magnetic field of 100 kHz-19.3 mT.

**II. Production of the magnetic field**

*1. General principle*

The general principle of our instrument is schematized in Fig. 1a. In a first approximation, an electromagnet can be modelled with its inductance $L_0$ and stray capacitance $C_0$. The main contribution to this stray capacitance is the capacitive coupling between the windings. However, other factors such as the influence of nearby conductors and the dielectric properties of the core must also be taken into account to explain its value [8]. As a consequence, the impedance of an electromagnet in this simple model is given by

$$Z_0 = \frac{iL_0\omega}{1 - L_0 C_0 \omega^2} \qquad (1)$$

where $\omega$ is the pulsation of the alternative signal. Below (respectively above) the self-resonant pulsation $\omega_0 = \frac{1}{\sqrt{L_0 C_0}}$, the major part of the current passes through the inductance (resp. capacitance). An electromagnet should therefore be used at $\omega < \omega_0$ for an efficient use of the power to produce a magnetic field. If a capacitance $C_1$ is placed in series with the electromagnet (see Fig. 1a), the impedance of the circuit becomes

$$Z = -i\left(\frac{1}{\omega C_1} + \frac{L_0 \omega}{\omega^2 L_0 C_0 - 1}\right) \qquad (2)$$

From this equation, it can easily be observed that, at the condition that $\omega < \omega_0$, the impedance of the circuit can in principle be brought to zero by choosing

$$C_1 = \frac{1-\omega^2 L_0 C_0}{L_0 \omega^2} \quad (3)$$

Actually, there is a residual impedance at the resonance, which can be calculated by taking in consideration the resistance of the circuit wires and a more accurate description of the electromagnet, i.e. a parallel RLC model (see Fig. 1b). Assuming that $C_1$ is still chosen to cancel the imaginary part of the impedance, the residual impedance at the resonance is then given by

$$Z = R_w + \frac{R_0 L_0^2 \omega^2}{R_0^2\left(1-\left(\omega/\omega_0\right)^2\right)^2 + L_0^2 \omega^2} \quad (4)$$

where $R_w$ is the resistance of the circuit wires and $R_0$ the parallel resistance of the electromagnet. This function, plotted in Figure 3c, displays a maximum at the self-resonant pulsation $\omega_0$ of the coil. As a conclusion, the guideline for the design of an efficient high-frequency electromagnet is to conceive it with a self-resonance much greater than the maximum working frequency and to minimize its impedance through the adjustable capacitance $C_1$. When $\omega << \omega_0$ and considering the fact that $R >> L_0 \omega$ in the experimental relevant cases (see below), an approximate expression for the residual impedance at the resonance is

$$Z \approx R_w + \frac{L_0^2 \omega^2}{R_0}. \quad (5)$$

Thus at low frequencies, the residual impedance increases with the square of the frequency.

## 2. Materials and optimization of the electromagnet

The electromagnet is simply composed of commercial I-shaped Ni-Zn ferrites used for power transformer (Epsos, material N27), mounted with a gap of 1.1 cm (see Fig. 2a). The voltage amplifier used is a HSA 4201 (current ± 2.8A, voltage ±150 V, frequency 500 kHz). Given these specifications, the electromagnet has been optimized for a maximum working frequency of 500 kHz, with the constraint that its impedance at this frequency should not exceed 50 Ω. The first point needed to be optimized for a high-frequency electromagnet is the nature of the wires used. Indeed, the impedance of standard copper wires increased dramatically at high frequency due to skin effects. Moreover, their impedance increases when a large current is flowing through them due to proximity effects. These classical problems are generally solved using Litz wires [9]. An optimal Litz wire should be weaved rather than twisted. For our experiments, we used a custom-made Litz wire (24×10×0.05, Connect systemes, France). It is composed of 24 brands weaved together, each brand being composed of 10 twisted copper wires with a diameter of 0.05 mm. As will be more evident below, the use of this wire was crucial to keep as low as possible the impedance of the electromagnet.

Several other parameters must be taken into account to decrease the impedance of the electromagnet. First, the stray capacitance of the windings increases with the number of turns of the coil. Thus, this number must be chosen so that the self-resonance of the electromagnet is much higher than the working frequency. Moreover, winding directly on the ferrite increases the stray capacitance of the electromagnet and creates hot spots and sparks when a large current is driven through the wires. To avoid this, the wire is winded on a 1 mm-thick plastic wrap for transformers. It has also been observed that the remaining impedance is decreased when (i) the wire is winded on the same side than the gap rather than on one of the closure-flux branches of the electromagnet, and (ii) a single layer of wire is used. Finally, for a given magnetic field / current ratio, it is better to use two identical coils with a given number of turns placed in parallel on each side of the gap rather than using a single coil with the same number of turns. In the parallel configuration, the current in each coil is divided by two and the total number of turns doubled, so the magnetic field / current ratio is unchanged, but the remaining impedance is again slightly lowered. Following these rules, for a working frequency of 500 kHz and a gap $e$ of 1.1 cm, the optimal number of turns is around 20 on each coil. A picture of the optimized electromagnet is shown in Fig. 2a. With this geometry, the magnetic field / current ratio is 1.93 mT/A, value in good agreement with the classical equation for electromagnets $B \approx \mu_0 NI/e$, where $N$ is the number of turns and $I$ the current in the coil.

Fig. 3a shows the electrical characterisation of the stand-alone electromagnet when a low current is driven through it. A function generator is used as a power supply. The frequency dependence of the impedance can be fitted by a simple LC model. $C_0$ = 3.5 pF and $L_0$ = 0.1 mH are deduced from the fit of these data using Equ. (1), leading to a self-resonance frequency $f_0$ = 8.5 MHz. In Fig. 3b and 3c, the characterisation of the electromagnet in series with the adjustable capacitance $C_1$ is shown. The optimised values of $C_1$ to minimize the value of the residual impedance are plotted as a function of frequency. The evolution of $C_1$ as a function of the frequency is in agreement with the expected dependence calculated with Equ. (3), using the

$C_0$ and $L_0$ values previously determined. The evolution of the residual impedance follows a square power law of the frequency, as expected from Equ. (5). Its fitting gives $L_0 = 0.1$ mH, $R = 4000$ Ω and $R_w = 0.5$ Ω. For our working frequency range, i.e. between 0 and 500 kHz, the impedance is kept below 12 Ω, that is a very low value.

To get the large currents necessary to produce the magnetic field, the function generator is combined with the voltage amplifier. We built an adjustable capacitance $C_1$ able to support large current in a broad range of frequencies by replacing the capacitors of a standard decade box (Centrad, DC05) by high-voltage ceramic disk capacitors (Panasonic, MDU series; Murata, DE series; Epcos, B81123 series). Finally the impedance of the circuit is characterized by measuring the voltage using the monitor output of the amplifier, and the current by using an ac current probe (Tektronix, P6021). Interestingly, the residual impedance weakly depends on the current driven in the circuit for working frequencies between 20 kHz and 500 kHz (see next paragraph for quantitative data). Thus there is a weak influence of proximity and skin effects on the electromagnet impedance, as expected with the use of the custom-made Litz wire. Given the magnetic field / current ratio of the electromagnet and its impedance, a magnetic field of 4.8 mT is generated at 100 kHz and 500 kHz with an output power of 3.4 W and 75 W, respectively.

### 3. Use of a resonant transformer

Given the very weak impedance of the electromagnet (0.56 Ω at 100 kHz), the limiting factor to further increase the magnetic field below 100 kHz is the current delivered by the voltage amplifier (limited to 3 A). To increase this value, a resonant transformer is added between the amplifier and the electromagnet. An adjustable capacitance $C_2$ similar to $C_1$ is used to bring the transformer at resonance (see Fig. 1c). The transformer is composed of the same material as the electromagnet and has 20 turns of Litz wire at the primary and 4 turns at the secondary. To bring the complete circuit at resonance the switch $K$ is first closed to isolate the transformer and $C_2$ is adjusted. Then the switch $K$ is opened to connect the two parts of the circuit and $C_1$ is adjusted. The ratio between the power in the primary and the power in the secondary at 100 kHz is 93 %. The evolution of the impedance in the secondary circuit as a function of the magnetic field produced is plotted on Fig. 3c at 100 kHz. A slight increase of the impedance (from 0.45 to 0.65 Ω) is observed when the field is increased from 0 to 19 mT. As a conclusion, using the resonant transformer a magnetic field of 19.3 mT is generated at 100 kHz with an output power of 65 W. When the output current of the voltage amplifier reaches its limit of 3 A, a maximum magnetic field of 29 mT is obtained.

**III. Hyperthermia measurements**

We then use this set-up to characterize the hyperthermia properties of Fe nanoparticles. Heating power is determined by measuring the temperature rise of a colloidal solution of magnetic nanoparticles. This colloidal solution can be either (i) put directly inside a 1 cm in diameter Teflon cylinder placed into the gap or (ii) first sealed into a glass tube under Argon atmosphere (Avitec, France) and then immerged in water in the Teflon cylinder. This packaging is necessary when using air-sensitive nanoparticles. The temperature is measured using an optical probe, unsensitive to alternative magnetic field (Neoptics, Reflex 4). The evolution of the ambient temperature and the self-heating of the electromagnet may be a source of artefacts, especially when highly diluted colloidal solutions or samples with a weak heating power are measured. To avoid this, a differential measurement is used. A blank sample similar to the one being measured but without nanoparticle is placed in a second calorimeter inside the gap. Its temperature is measured with a second optical probe (see Fig. 2a). The difference of temperature between the two samples can be directly attributed to the heating of the nanoparticles.

Fig. 4a shows the temperature evolution of two namely identical blank samples, simply composed of 4 ml of distilled water and measured for an output power of 71 W. The temperature rise is only due to the heating of the electromagnet-coil system. After a delay of about 2 mn a linear increase of temperature with time is observed. From this linear slope the amount of power transferred from the electromagnet to the calorimeter is deduced. Similar experiments for various output power values have been performed and are summarized in Fig. 4b. From this plot, we estimate that approximately 0.02 % of the output power in transferred to the calorimeter. In Fig. 4a, the evolution of the differential temperature is also plotted, illustrating that the use of differential measurement improves the sensitivity of the experiment by about an order of magnitude. The sensitivity $S$ of the experiment can as a consequence be expressed as $S = 2\times10^{-5}.O$, where $O$ is the output power of the amplifier. For instance, at 100 kHz and 19 mT, the output power of the amplifier is 65 W, which means that a heating power above 1.5 mW can be measured. Moreover, as will be shown below, because of the large difference of rising time between the heating power due to nanoparticles and the one due to the electromagnet heating, heating values below this limit can be distinguished.

Finally, we present measurements performed on a colloidal solution of 7.5 nm in diameter Fe nanoparticles elaborated by organometallic chemistry [10]. These nanoparticles are sensitive to oxidation so they are first embedded in a glass container under inert atmosphere. The container is filled with 19 mg of nanoparticles, 129 mg of organic ligands, and 800 mg of solvent (Mesitylene). The concentration in nanoparticles is about 0.3 % vol. The blank sample consisting of a glass container filled with 929 mg of Mesitylene is measured simultaneously. In Fig. 5a and b, the experiments performed at 15.5 mT at a frequency of 20 and 100 kHz are shown. In the first (second) case, the heating due to the nanoparticle is similar (much higher) than the heating due to the electromagnet. In both cases, the temperature rise is nearly immediate in the nanoparticle sample, while the temperature in the reference container only increases after 2 min. The heating induced by the nanoparticles is well evidenced by the differential measurement even in the first case: a clear sign for this is the decrease of temperature in the differential measurement as soon as the magnetic field is switched off, even if the raw temperature measurement on the sample with nanoparticle still increases. Fig 5c displays the dependency versus the magnetic field at a single frequency of 100 kHz. The evolution of the heating power of the nanoparticles as a function of the frequency under a magnetic field of 15.5 mT is shown in Fig. 5d. The variations of heating power with frequency and magnetic field follow a power law with an exponent 1.2 and 1.8, respectively. The magnetic field dependence is typical of a Néel relaxation process. The fact that the heating power increases more than linearly with frequency means that the resonant frequency of the nanoparticles is above the maximum frequency we applied. The nanoparticle diameter should thus be increased to optimize the heating power value. More details on these experiments and their discussion will be published in a forthcoming article. These experiments illustrate that even low heating power on a highly diluted colloidal solution can be accurately measured on the bench. Moreover, the knowledge of the frequency-dependence of the heating power is a useful tool to optimize hyperthermia properties.

**IV. Conclusion**

We have described an apparatus suitable for hyperthermia measurements on colloidal solutions of magnetic nanoparticles. It is frequency-adjustable and produces a large magnetic field without the requirement of a large output power. Many experimental details on the construction of this low-cost set-up have been provided so it could be easily reproduced in other laboratories working on hyperthermia. The measurements of hyperthermia at various frequencies and magnetic field allow one to go further into the mechanisms of hyperthermia and to correlate the heating power of the nanoparticles to their magnetic properties.

**Acknowledgements :**
We acknowledge O. Brouat, A. Le Masson, C. Lemaignan and J. Delbut for their experimental work on the optimisation of the setup, C. Crouzet for her precious help on the capacitor box, and J. Moreau for the machining of mechanical pieces.

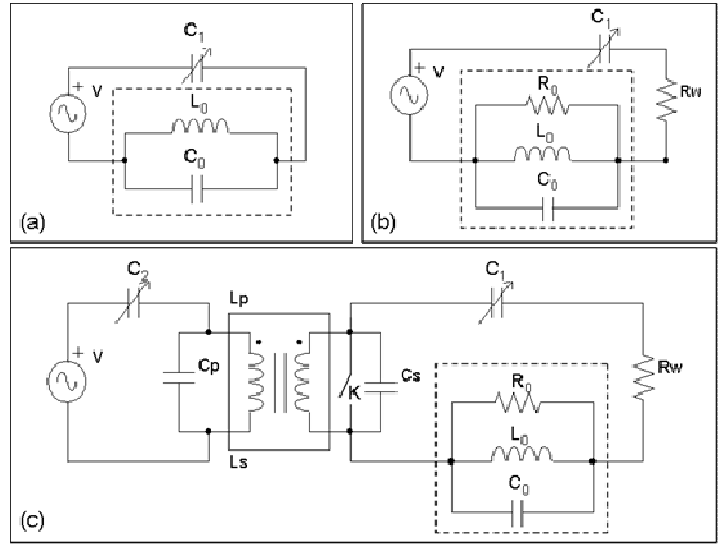

Figure 1: Schematic of our setup. The dashed box represents the electromagnet itself. (a) The electromagnet is described by an inductor/capacitor model. $C_1$ is a variable capacitor. (b) The electromagnet is modelled by a parallel RLC circuit and the resistance of the wires $R_w$ is taken into account (c) Complete schematic of the optimized transformer/electromagnet setup.

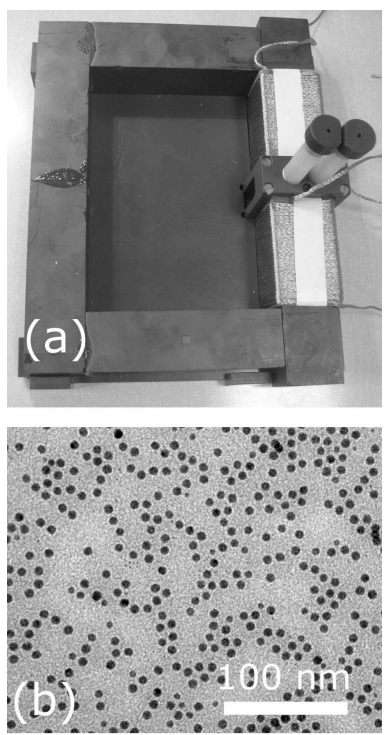

Figure 2: (a) Picture of the optimized electromagnet. (b) Transmission electron microscopy micrograph of the iron nanoparticles synthesized by organometallic chemistry.

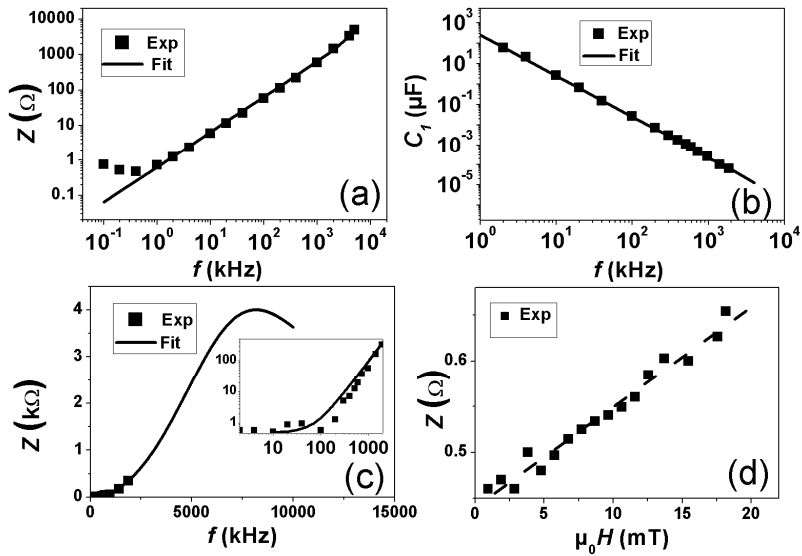

Figure 3 : Electrical characterisation of the optimized electromagnet. All the fits uses the same values $L_0 = 0.1$ mH, $C_0 = 3.5$ pF, R = 4000 Ω and Rw = 0.5 Ω. a) Total impedance of the stand-alone electromagnet as a function of the working frequency, and fit using Equ. (1). (b) Evolution of the $C_1$ value necessary to bring the circuit at resonance as a function of frequency, and fit using Equ. (3). (c) Evolution of the remaining impedance as a function of frequency and fit using Equ. (5) (inset) enlarged view of the same data. (d) Evolution of the remaining impedance as a function of the magnetic field for experiments performed at 100 kHz. The dashed line is a guide to the eye.

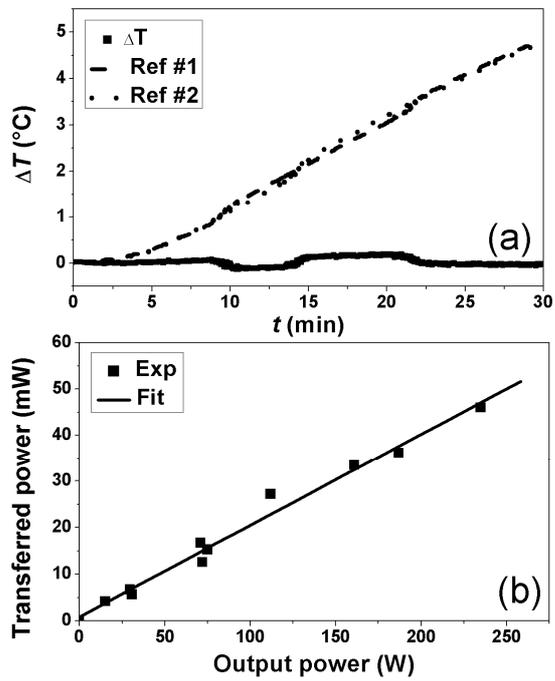

Figure 4 : (a) Temperature rise measured for 2 blank samples for an output power of the amplifier of 71 W, plotted alongside the difference between the two temperatures. (b) Power transferred to the calorimeter as a function of the output power of the amplifier.

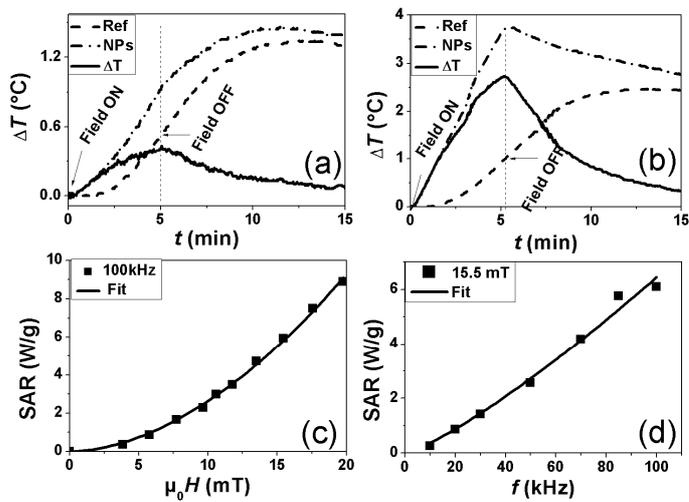

Figure 5: (a) Hyperthermia measurements performed at 15.5 mT-20 kHz on iron nanoparticles. The temperature of the reference sample (dot line), of the nanoparticle sample (dash-dot line) and the difference between both (plain line) are displayed. (b) Similar experiments performed at 15.5 mT-100 kHz. (c) Specific absorption rate of the iron nanoparticles as a function of magnetic field, fitted using a power law with an exponent 1.8. (d) Specific absorption rate of the iron nanoparticles as a function of frequency, fitted using a power law with an exponent 1.2.